\documentclass[aps,prl,twocolumn,superscriptaddress,groupedaddress]{revtex4}  

\usepackage{CJK}
\usepackage{graphicx}  
\usepackage{dcolumn}   
\usepackage{bm}        
\usepackage{amssymb}   
\usepackage{latexsym,bm}
\usepackage[tbtags]{amsmath}

\usepackage{color}

\usepackage{braket}

\begin{document}

\begin{CJK*}{GBK}{song} 

\widetext


\title{Universal scaling law of an origami paper spring}

\author{Bohua SUN}
 \altaffiliation[Also at ]{Physics Department, XYZ University.}

\affiliation{Institute of Mechanics and Technology \& School of Civil Engineering, Xi'an University of Architecture and Technology, Xi'an 710055, China\\
http://imt.xauat.edu.cn\\
email: sunbohua@xauat.edu.cn
}

\date{September 3, 2019 }


\email{sunbohua@xauat.edu.cn}

\begin{abstract}
This letter solves an open question of origami paper spring raised by Yoneda et al.(2019). By using both dimensional analysis and data fitting, an universal scaling law of a paper spring is formulated. The scaling law shows that origami spring force obeys power square law of spring extension, however strong nonlinear to the total twist angle. The study has also successfully generalized the scaling law from the Poisson ratio 0.3 to an arbitrary Poisson's ratio with the help of dimensional analysis.
\end{abstract}



\maketitle


The mechanics of elastic origami structures has become a popular topic in the field of physics (Faber \cite{4}, Hull \cite{5},  Lechenault and Adda-Bedia \cite{7}, Haas and Wootton \cite{8},  Ren \emph{et al.} \cite{12}, Deboeuf \emph{et al.} \cite{13}, Cheng and Zhang \cite{18}). In contrast to the classic rigid origami, the complete understanding of the geometric mechanics of the elastic origami is still a challenge subject.
\begin{figure}[h]
\centerline{\includegraphics[scale=0.03]{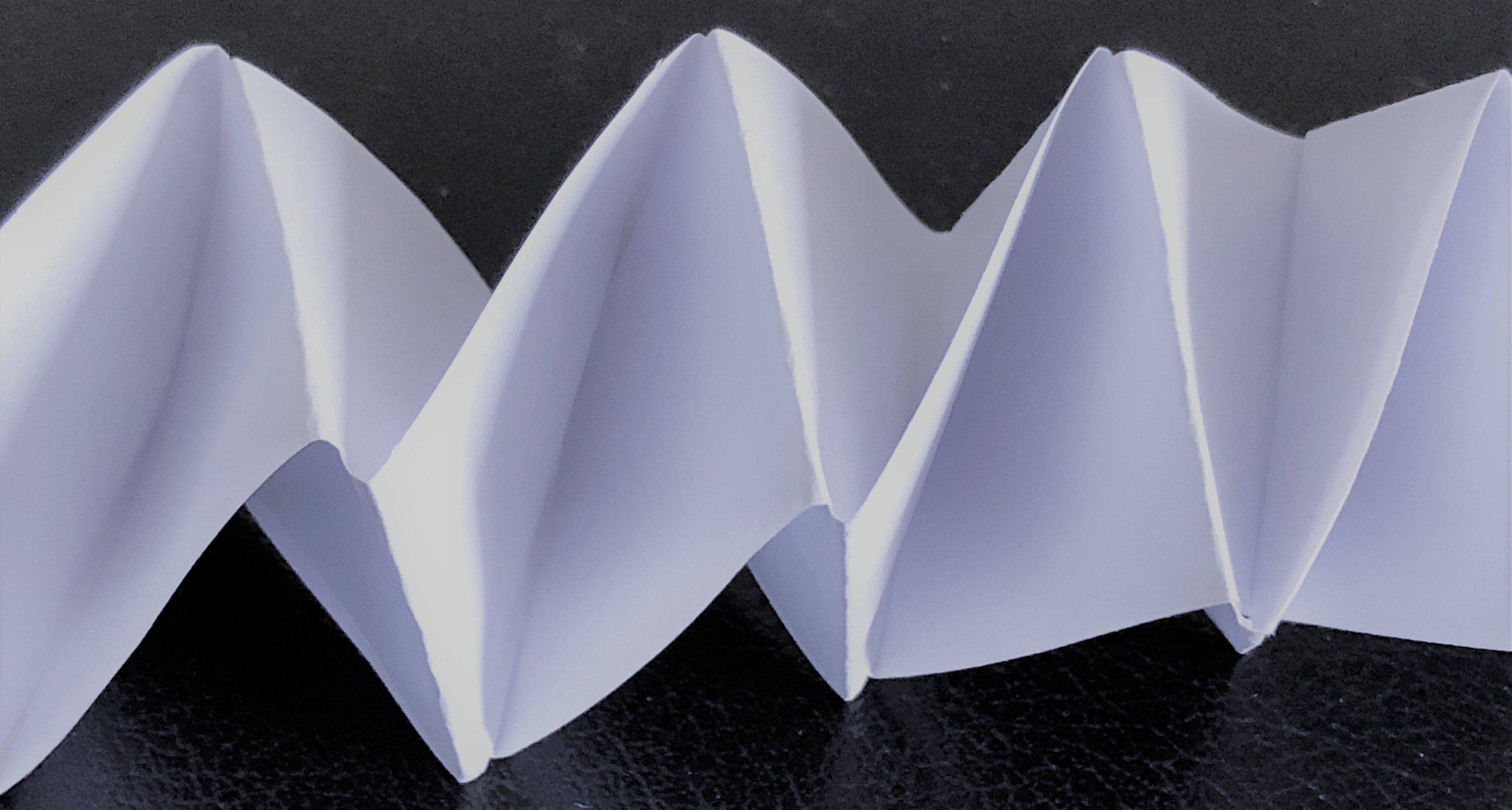}}
\caption{\label{fig-spring} Making of a paper spring. (1) Begin with two long
strips of paper of equal sizes, glue them together such that they form a
right angle. (2) Fold the lower strip over the top one, then (3) fold the
strip that just came to the lower layer over the one on top. Continue
this process (the lower strip always folds over the top one), until the
last fold at which the entire lengths of the strips are folded up into a
small square. The process is completed by gluing the last flap \cite{p-spring}. }
\end{figure}

Yoneda \emph{et al.} \cite{p-spring} recently reported a remarkable result on a paper spring contracted by interfolding and gluing two long strips of paper of equal sizes (shown in Figure \ref{fig-spring}). Through systematic study, they have revealed the strong geometric coupling between stretch and twist, as well as the increasing stiffness against a global bending deformation during unfolding.

However, Yoneda \emph{et al.} \cite{p-spring} also left an open and intriguing question as follows:
\begin{quote}
For large extension, an empirical spring force scaling law is proposed:
\begin{equation}\label{eq1}
  F=\frac{Et^{3-\gamma}}{a^{1-\gamma}}f(\frac{z}{aN}),
\end{equation}
the question is that whether the exponent $\gamma=3/4$, as well as the unknown scaling function $f(x)$, are universal for paper springs with different geometries, where Young modulus $E$, plate thickness $t$ and side $a$, paper spring extension $z$, plate periodicity number $N$ and spring force $F$.
\end{quote}

To answer this question, we must find out how to get Eq.(\ref{eq1}), since Yoneda \emph{et al.} \cite{p-spring} does not give any information on the formulation of Eq.(\ref{eq1}). 

It is clear that this problem has no exact solution, hence numerical solutions and experimental methods have to be anticipated. However, the numerical and experimental results can only be obtained for a specific case, it would be very difficulty to predict the general scaling trends on the problem by using limited numerical results. To find the general scaling law, it would be a natural attempts to take an alternative way - dimensional analysis (Bridgman \cite{bridgman1922}, Sun \cite{sun,sun2017,sun2018-a,sun2018-b}).

As we know, any physical relationship can be expressed in a dimensionless form. The implication of this statement is that all of the fundamental equations of physics, as well as all approximations of these equations and, for that matter, all functional relationships between these variables, must be invariant under a dilation of the dimensions of the variables. This is because the variables are subject to measurement by an observer in terms of units that are selected at the arbitrary discretion of the observer. It is clear that a physical event cannot depend on the particular ruler, which is used to measure space, the clock is used for the time, and the scale is used to measure mass, or any other standard of measure that might be required, depending on the dimensions that appear in the problem. This principle is the basis for a powerful method of reduction, which is called dimensional analysis and is useful for the investigation of complicated problems. Often, dimensional analysis is conducted without any explicit consideration for the actual equations that may govern a physical phenomenon (Bridgman \cite{bridgman1922}, Sun \cite{sun,sun2017,sun2018-a,sun2018-b}).

From a physics point of view, dimensional analysis is an universal method, which can, of course, be used for the study of origami paper spring. We first prove that the Eq.\ref{eq1} is universal from dimensional perspectives, then determining $\gamma$ and $f(x)$ by data fitting. Why we use the dimensional analysis, it is because that there is consensus the universality can be verified by the dimensional analysis. It means the relation in Eq.(\ref{eq1}) is universal if it could be formulated by the dimensional methods.

For the deformation of origami paper spring, there are 7 quantities, namely $F,\,z,\,E,,\nu,\,\,a,\,t,\,N$, where $\nu$ is the Poisson's ratio. The quantity dimensions are listed in the following table \ref{table}:
\begin{table}[h]
\caption{Dimensions of physical quantity}\label{table}
\begin{tabular}{c|c}
\hline
Variables &  Dimensions \\
\hline
$F$ &  $\mathrm{MLT^{-2}}$  \\
\hline
$z$ &  $\mathrm{L}$  \\
\hline
$E$ &  $\mathrm{ML^{-1}T^{-2}}$  \\
\hline
$\nu$ &  1  \\
\hline
$t$ &  L  \\
\hline
$a$ &  L  \\
\hline
$N$ &  1  \\
\hline
\end{tabular}
\\
\footnotesize The dimensional basis used is length (L), mass (M) and time (T).
\end{table}

The spring force $F$ can be expressed as the function of quantities $z,E,\nu,a,t,N$, namely,
\begin{equation}\label{eqf-1}
  F=F(z,E,\nu,a,t,N).
\end{equation}
In the above relation, there are 7 quantities, two of them are dimensionless. Since only 3 dimensional basis $L,\,M,\,T$ are used, so according to dimensional analysis Bridgman \cite{bridgman1922}, Sun \cite{sun2017}, Eq. (\ref{eqf-1}) produces two dimensionless quantities $\Pi$, the first one is:
\begin{align}\label{eqx1}
   \Pi_1&=\frac{z}{aN},
\end{align}
From experimental observation, it is found the paper spring bending stiffness plays a domination role rather than the Young modulus $E$, therefore, it would be more natural to rewrite the expression Eq. (\ref{eqf-1}) in following format:
 \begin{equation}\label{eqf-2}
  F=F(\frac{z}{aN},B,a,t),
\end{equation}
where the paper spring bending stiffness $B=\frac{Et^3}{12(1-\nu^2)}$ with dimensions $ML^{2}T^{-2}$.

The second dimensionless quantities $\Pi$ is in the form:
\begin{align}\label{eqx2}
  \Pi_2&=FB^\alpha a^\beta t^\omega,
\end{align}
where $\alpha,\,\beta$ and $\omega$ are constant, which must satisfy following dimensions condition:
\begin{align}\label{eqx3}
  \mathrm{dim}(\Pi_2)&=M^0L^0T^0 \\
  &=MLT^{-2}(ML^{2}T^{-2})^\alpha L^\beta L^\omega\\
  &=M^{1+\alpha}L^{1+2\alpha +\beta+\omega}T^{-2-2\alpha}.
\end{align}
From the condition, we have
\begin{align}\label{eqx4}
  M: & 1+\alpha=0,  \\
  L: & 1+2\alpha +\beta+\omega=0,\\
  T: & -2-2\alpha=0,
\end{align}
which gives $\alpha=-1$, and $\beta+\omega=1$. With these solution, we have
\begin{align}\label{eqx2-1}
  \Pi_2&=\frac{F}{B} t(\frac{a}{t})^\beta,
\end{align}
Based on the Buckingham $\Pi$ theorem (Bridgman\cite{bridgman1922}, Sun \cite{sun,sun2017}), the scaling law of the paper spring must be expressed as follows: $\Pi_2=\bar{f}(\Pi_1)$, namely
\begin{equation}\label{xx}
  \frac{F}{B} t(\frac{a}{t})^\beta= \bar{f}(\frac{z}{aN}).
\end{equation}
Hence,
\begin{align}
  F&=B\frac{1}{t}(\frac{t}{a})^\beta f(\frac{z}{aN})\nonumber \\
  &=\frac{Et^3}{12(1-\nu^2)}\frac{1}{t}(\frac{t}{a})^\beta \bar{f}(\frac{z}{aN})
\end{align}
Introducing a new constant by $\beta=1-\gamma$, the above relation becomes
\begin{align}\label{xyz}
  F=E\frac{t^{3-\gamma}}{a^{1-\gamma}}f(\frac{z}{aN}).
\end{align}
where $f(x)=\bar{f}(x) /[12(1-\nu^2)]$, and $\gamma$ is a constant to be determined.

It is clear that Eq.(\ref{xyz}) is same as to Eq.(\ref{eq1}). Haven dimensions verification, we can say that Eq.(\ref{eq1}) is universal.

By data fitting, Yoneda \emph{et al.} \cite{p-spring} determined $\gamma=3/4$ but could not figured out the scaling function $f(x)$.

\begin{figure}[h]
\centerline{\includegraphics[scale=0.5]{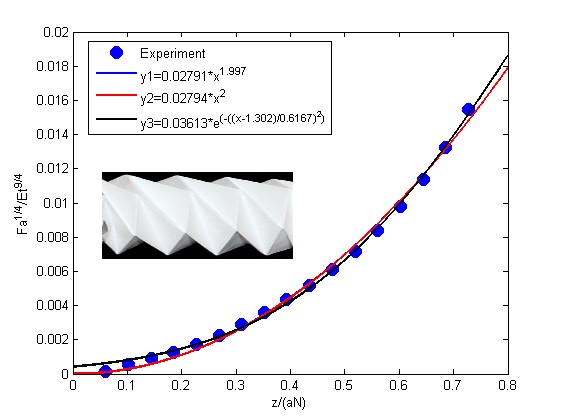}}
\caption{\label{fig-1} Force vs extension curves in re-scaled form proposed in Eq.(\ref{eq1}), power exponent is close to $2$.}
\end{figure}

To determine the function $f(x)$, using data taken from Figure 7(c) in Yoneda \emph{et al.}\cite{p-spring} and numerical data fitting, three scaling functions $f(x)$ are generated and shown in Figure \ref{fig-1}.

The data fitting reveals that the scaling function obeys law that is close to a power law, namely
\begin{equation}\label{f}
  f(x) \sim x^2.
\end{equation}
Therefore, we have the universal relation for paper springs with different geometries as follows:
\begin{equation}\label{eq2}
\begin{split}
  F &=0.02794 \frac{Et^{9/4}}{a^{1/4}}(\frac{z}{aN})^2. 
  \end{split}
\end{equation}
Since $\frac{z}{aN}$ can be expressed in terms of the total twist angle $\Phi(z)$ as follows \cite{p-spring}:
\begin{equation}\label{z}
  \frac{z}{aN}=\frac{4}{\sqrt{1+\sin(\frac{\Phi}{4N})}}.
\end{equation}
Hence, the spring force $F$ also can be expressed in terms of total twist angle $\Phi(z)$ as follows:
\begin{equation}\label{eq3}
  F =0.44704\frac{Et^{9/4}}{a^{1/4}}\frac{1}{1+\sin(\frac{\Phi}{4N})}.
\end{equation}
Noticed all data and analysis in Yoneda \emph{et al.} \cite{p-spring} were done for the material with Poisson's ratio $\nu=0.3$, the question is that how to extend the above scaling law to other materials with different Poisson's ratio.

Since we have proved that the scaling law Eq.(\ref{eq1}) is universal, therefore, we can easily generalize all scaling law to the materials with an arbitrary Poisson's ratio as follows:
\begin{equation}\label{final-1}
  F =0.02543 \frac{E}{1-\nu^2}\frac{t^{9/4}}{a^{1/4}}(\frac{z}{aN})^2,
\end{equation}
and origami paper spring stiffness
\begin{equation}\label{final-2}
  K(z)=\frac{dF}{dz} =\frac{0.05086}{N^2} \frac{E}{1-\nu^2}(\frac{t}{a})^{9/4}z.
\end{equation}
These relation is obtained naturally without to do additional works. Eq.(\ref{final-1}) is plotted in Figure \ref{fig-3} below.
\begin{figure}[h]
\centerline{\includegraphics[scale=0.3]{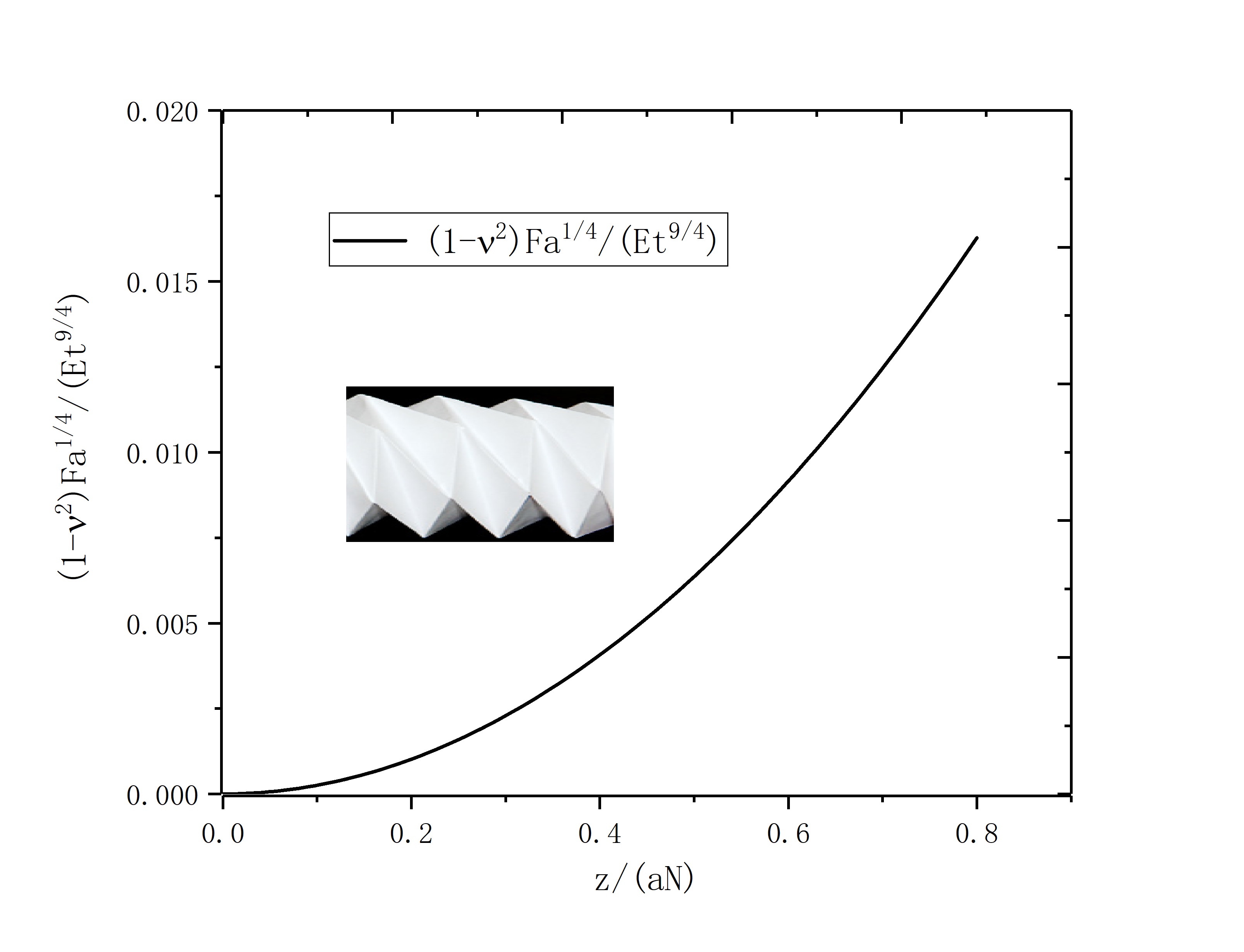}}
\caption{\label{fig-3} Force vs extension curves in re-scaled form proposed in Eq.(\ref{final-1}) for arbitrary Poisson's ratio.}
\end{figure}

The spring force $F$ vs twist angle $\Phi$ is given by
\begin{equation}\label{final-3}
  F =0.4068\frac{E}{1-\nu^2}\frac{t^{9/4}}{a^{1/4}}\frac{1}{1+\sin(\frac{\Phi}{4N})},
\end{equation}
and force-twist stiffness is
\begin{equation}\label{final-4}
  \frac{dF}{d\Phi}=-\frac{0.1017}{N}\frac{E}{1-\nu^2}\frac{t^{9/4}}{a^{1/4}}\frac{\cos(\frac{\Phi}{4N})}{\left[1+\sin(\frac{\Phi}{4N})\right]^2}.
\end{equation}
Both relations in Eq.(\ref{final-1}) and Eq.(\ref{final-3}) reveal that the origami paper spring is a strong nonlinear for both extension and twist angle.

In conclusion, this Letter has solved an open question raised by Yoneda \emph{et al.} \cite{p-spring}, and an universal scaling laws of origami paper spring has been formulated. Even beyond the scope of Yoneda \emph{et al.}'s work \cite{p-spring}, we have successfully generalized the scaling law from the Poisson's ratio 0.3 to an arbitrary Poisson's ratio with the help of dimensional analysis.

\textbf{Acknowledgement}: The author appreciates the financial supports from Xi'an University of Architecture and Technology and Mr Zhe Liu for the preparation of Figure \ref{fig-1}.



\end{CJK*}

\end{document}